\documentclass[conference]{IEEEtran}
\IEEEoverridecommandlockouts
\usepackage{cite}
\usepackage{amsmath,amssymb,amsfonts}
\usepackage{algorithmic}
\usepackage{graphicx}
\usepackage{textcomp}
\usepackage{xcolor}
\usepackage{booktabs}
\def\BibTeX{{\rm B\kern-.05em{\sc i\kern-.025em b}\kern-.08em
    T\kern-.1667em\lower.7ex\hbox{E}\kern-.125emX}}
\begin{document}

\title{EPIC: Generative AI Platform for Accelerating HPC Operational Data Analytics
\thanks{Identify applicable funding agency here. If none, delete this.}
}

\author{\IEEEauthorblockN{Ahmad Maroof Karimi, Woong Shin, Jesse Hines, Tirthankar Ghosal, Naw Safrin Sattar, Feiyi Wang}
\IEEEauthorblockA{\textit{National Center for Computational Sciences} \\
\textit{Oak Ridge National Laboratory}\\
Oak Ridge, USA \\
\{karimiahmad, shinw, hinesjr, ghosalt, sattarn, fwang2\}@ornl.gov}
}
\maketitle
\begin{abstract}
    We present EPIC, an AI-driven platform designed to augment operational data analytics. EPIC employs a hierarchical multi-agent architecture where a top-level large language model provides query processing, reasoning and synthesis capabilities. These capabilities orchestrate three specialized low-level agents for information retrieval, descriptive analytics, and predictive analytics. 
    This architecture enables EPIC to perform HPC operational analytics on multi-modal data, including text, images, and tabular formats, dynamically and iteratively.  EPIC addresses the limitations of existing HPC operational analytics  approaches, which rely on static methods that struggle to adapt to evolving analytics tasks and stakeholder demands. 
    
    Through extensive evaluations on the Frontier HPC system, we demonstrate that EPIC effectively handles complex queries. Using descriptive analytics as a use case, fine-tuned smaller models outperform large state-of-the-art foundation models, achieving up to 26\% higher accuracy. Additionally, we achieved 19x savings in LLM operational costs compared to proprietary solutions by employing a hybrid approach that combines large foundational models with fine-tuned local open-weight models.
\end{abstract}


\section{Introduction}
\label{sec:intro}

Extracting actionable insights from HPC operations has become increasingly important in addressing complex operational challenges rising in the exascale era. 
Operational data analytics (ODA) systems are designed to derive these insights from multi-terabyte data streams originating from modern HPC systems, and play a critical role in optimizing performance, energy usage, reliability, resource utilization, and procurement planning.
Extensive research has been conducted in ODA using telemetry data gathered during HPC operations to study system behavior under various settings and workloads~\cite{netti2019dcdb,netti2020wintermute,patel2020does,shin2021revealing,ott:2020,Bautista:2019,Bourassa2019,shoga2018,wilde20144,icdcs_olcf_24,hpc_workload_trace}. 

However, the efficacy of conventional ODA techniques is limited when addressing the rapidly evolving needs and increasingly diverse set of stakeholders, as they do not provide dynamic, on-demand, and iterative analytical environments~\cite{NETTI2022102950}. ODA tasks on large datasets are time-consuming and require expertise in data engineering, data science and machine learning. Furthermore, conventional monitoring dashboards offer a static, pre-defined array of visualization charts \cite{dashboard_1, dashboard_2}, further constraining their impact under the on-demand and evolving needs of a diverse stakeholder. 

The above limitations point to the need for a more flexible, intelligent system capable of adapting to diverse and evolving analysis needs.
To address these limitations and enhance insights derived from HPC operational data, we present the design and implementation of EPIC, an AI-driven platform that leverages the new emerging capabilities of generative AI foundational models \cite{hong2023metagpt, huang2024understanding, zou2024cooperative} into the context of HPC operational analytics. EPIC aims to augment various data driven tasks such as descriptive analytics and predictive analytics by automating the process of reasoning and interacting with high-dimensional multi-modal HPC operational data and synthesizing the results into meaningful insights.

EPIC achieves this goal using a hierarchical multi-agent architecture. This architecture combines generic reasoning and synthesis capabilities of large foundational models with specialized low-level models. These low-level, fine-tuned, open-weight models are integrated into the HPC operational data infrastructure providing access to 1) an unstructured knowledge-base of text, tabular and image data, 2) analytical query into large, structured telemetry data, and 3) seamless interaction with predictive machine learning models trained on past data.
Also, to avoid the high-cost of maintainability and operation of large language models, EPIC leverages a hybrid of powerful proprietary foundational models and local open-source models. As a result, EPIC provides a highly modular yet extensive platform that can easily integrate diverse and heterogeneous operational data analytics capabilities while maintaining cost effectiveness.

Our contributions are as follows:
\begin{enumerate}
    \item We design an implementation that demonstrates the role and impact of large language models specialized for HPC operational analytics. The EPIC platform offers on-demand and iterative capabilities, providing a flexible and dynamic interface for exploratory analytics, and is designed to meet the continuously evolving analytical needs. 
    \item We present a hierarchical multi-agent architecture that facilitates the implementation of cost-effective specialization of generic AI capabilities for HPC operational analytics. EPIC's architecture introduces necessary abstractions to manage the underlying complexities of ODA tasks and provides a user-friendly interactive interface, thereby reducing the need for frequent reliance on data scientists and AI experts.
    \item We demonstrate the multifaceted analytical capabilities of the EPIC platform by conducting an extensive evaluation across a diverse range of HPC data-driven queries. We also facilitate the evaluation datasets along with the codebase\footnote{The codebase and data is shared as a part of reproduciblity initiative and AD/AE submission.}.
\end{enumerate}

The rest of the paper is structured as follows: Section~\ref{sec:background} provides a background on current practices in ODA and the applications of LLMs. Section~\ref{sec:design} explains the design and implementation of each component of the EPIC system, including a description of the evaluation pipeline for each module. Section ~\ref{sec:data} presents the EPIC software and the various datasets used in evaluating the models. Section~\ref{sec:evaluation} provides the evaluations of each module and highlights significant findings for each. Finally, Section~\ref{sec:conclusion} concludes the paper.

\begin{figure*}[t]
  \centering
  \includegraphics[width=18cm, height=6cm]{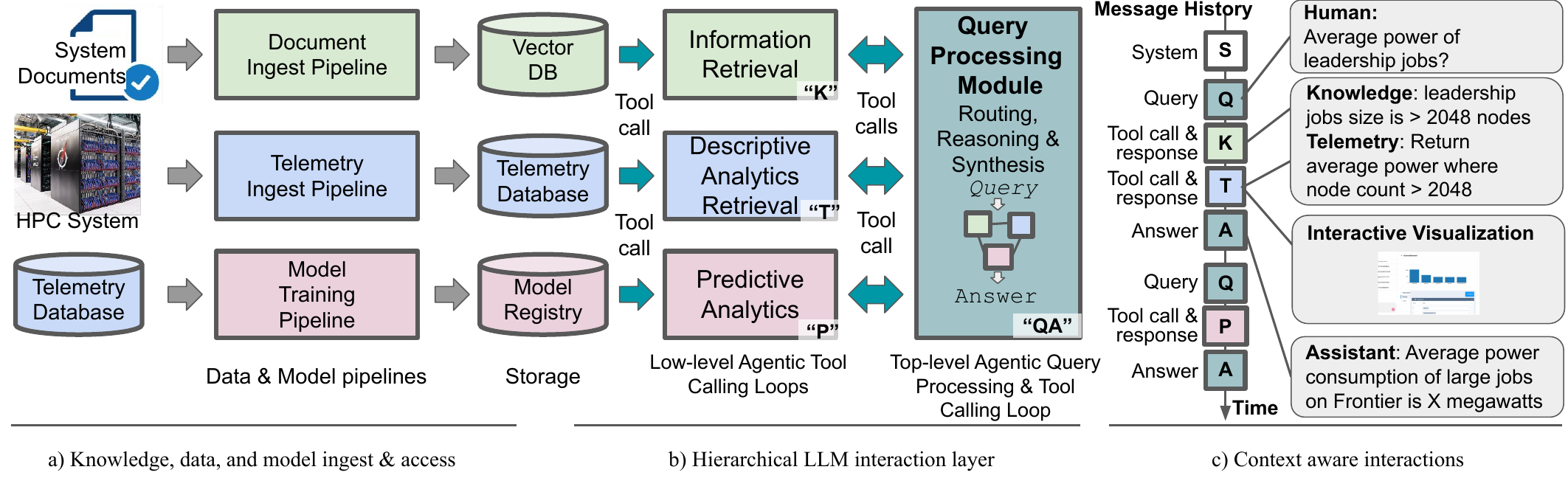}
  \caption{Architecture of the AI assisted data exploration system}
  \label{figs:architecture}
\end{figure*}

\section{Background and Motivation}
\label{sec:background}
This section outlines the current state of practice efforts in the domain of ODA and the utilization of LLMs and multi-tool agents to enhance analytics. We also demonstrate the applications of LLM and its utilization in various information retrieval and analytical activities.

\textbf{Operational Data Analytics and its challenges:}
Exascale HPC systems, mainly due to the increasing physical footprint caused by the diminishing return in performance per watt in the post-Moore era, the physical side of HPC systems introduces significant challenges such as Power, Energy and Reliability \cite{bergman:2008:exascalecomputing}. While fulfilling the necessary performance, HPC systems need to consider power, energy, cooling, and reliability, balancing many different KPIs all at the same time, thus being exposed to unprecedented cross-cutting complexity \cite{ref:4pillar:2014}.
 
Many HPC sites on the Top500 list are turning towards data as a key resource that can help address the challenges with deploying and operating these large-scale HPC systems \cite{ott:2020}. By establishing a holistic data collection campaign that targets the entire HPC data center, these HPC sites employ powerful data analytics infrastructures that can cope with the 3 Vs of Big Data (velocity, velocity, and variety) seen from these systems.
However, this introduces an inundation of data. Abundance of data enabled such comprehensive ODA infrastructures to introduce new opportunities for innovative data-driven decisions, however, there remain significant challenges in making impact with data. The skill-gap in HPC organizations lacking enough in-house data science expertise introduces bottlenecks in areas such as exploratory data analysis and predictive analysis of datasets \cite{shin:2024-oda}. 

The EPIC analytics platform can reduce this gap by facilitating HPC ODA tasks for users with varying degrees of data science expertise, thereby diminishing reliance on data science specialists.

\textbf{Current practices: Descriptive and Predictive Analytics:}
Netti et. al. \cite{netti2021} introduces a conceptual framework for HPC operational data analytics that can be applied to distinguish different capabilities of data analytics seen in HPC centers. The framework maps descriptive \& diagnostic (hindsight), and predictive \& prescriptive (foresight) data analytics \cite{lepenioti2020prescriptive} use cases in operational domains spanning across four different pillars~\cite{ref:4pillar:2014}.

In practice, the most frequent hindsight-driven analytics is descriptive, manifested in capabilities such as dashboards and interactive visualizations that can answer the question of “what happened?” \cite{ott:2020}. 
Also, in foresight, use cases in predictive analytics are frequently explored, reflecting the excitements in the potential of more advanced data science techniques manifesting in the form of various machine learning models trained from past data and being deployed to make predictions of future system behavior~\cite{sukhija_towards_2022, molan_semi-supervised_2022, shoukourian_forecasting_2020, bartolini_paving_2019} that can enhance existing dashboards or visual analytics systems.

However, such use cases suffer from the rigidity of these dashboards or analytics services both consuming long-development cycles \cite{shin:2024-oda} and being quite fixed to a certain use-case or a scenario~\cite{dashboard_1}. 
Meeting the analytics requirements of an HPC site is challenging because of continuously evolving inquiries. This inspired us to develop EPIC to incorporate dynamic, interactive methods for engaging with operational data to do exploratory data analytics and derive insights. 

\textbf{Applications of Large Language Models:}
Large Language Models (LLMs) have showcased impressive capabilities across diverse domains, exhibiting significant proficiency in reasoning, information processing, tool utilization, planning, and synthesis for various tasks including analytical jobs~\cite{hong2023metagpt, huang2024understanding, zou2024cooperative}. AI-driven multi-agent platforms further extend these capabilities by enabling collaboration among specialized agents that can interact, synthesize information and share intermediate results to solve complex problems\cite{zou2024cooperative}. 

In high-performance computing (HPC) and scientific research, open LLMs such as LLaMA series \cite{touvron2023llama}, Falcon \cite{almazrouei2023falcon}, Mistral \cite{jiang2023mistral}, BLOOM \cite{bigscience2022bloom}, etc. have been applied to automate code generation \cite{ValeroLara2023ComparingLA, kadosh2023scope}, parallelization \cite{10.1007/s10766-024-00778-9}, and workload orchestration \cite{Kadosh2023MonoCoderDC}.
Frameworks such as HyCE \cite{Miyashita2024LLMAH} further enhance LLM utility by grounding responses in real-time cluster state, improving the safety and reliability of LLM-driven system interactions in HPC environments.
Similarily, chatHPC \cite{Yin2024chatHPCEH} based on FORGE \cite{Yin2023FORGEPO} is another application of LLMs for High-Performance Computing knowledge discovery and question answering.


In system monitoring, LLMs fine-tuned with system logs and observability data are being applied to summarize HPC/cloud system failures \cite{10.1145/3643916.3644408}, detect anomalies \cite{Egersdoerfer2023EarlyEO}, and even suggest mitigation strategies \cite{Qi2023LogGPTEC}, laying the groundwork for self-healing infrastructure in scientific computing environments. All these works underscore how open LLMs are revolutionizing scientific computing by enhancing operational intelligence, energy-aware scheduling, and user accessibility across both HPC and cloud platforms.

Inspired by these advances, we are motivated to explore the application of LLMs in the context of HPC operational data analytics. We envision the EPIC platform assisting its users in navigating and analyzing heterogeneous HPC operational data, thereby solving the "last-mile problem" in ODA.

\section{EPIC Design and Implementation}
\label{sec:design}
In this section, we begin with a high-level, end-to-end description of the EPIC architecture. Subsequently, we delve into the design of each module. Then we present the design of the evaluation pipeline.

\subsection{Architecture}
\label{subsec:architecture}
 Figure~\ref{figs:architecture} illustrates the architecture of the system. Here, we primarily address the integration of 
 analytical tools driven by LLMs into HPC operational data analytics infrastructure by 1) identifying which LLMs work best for accuracy, efficiency, and performance and 2) by employing the right structures for workflow orchestration and develop a complex and dynamic \texttt{Query Processing Module}  and \texttt{top-level} model for navigating the users queries and its impact on the rest of the system.
\subsubsection{Framework}
\label{subsec:architecture:framework}
With the data and model access layer (Fig.~\ref{figs:architecture}-a) and the hierarchical LLM interaction layer (Fig.~\ref{figs:architecture}-b), the architecture provides a composable way to 
accommodate various low-level capabilities of HPC operational data analytics into a single intelligent system. Each low-level capability tool
involves accessing the influx of various data inputs from the HPC operational environment (Fig.~\ref{figs:architecture}-a). Such capabilities include the retrieval of domain knowledge and interactions with the telemetry database and predictive models, which are then exposed to the LLM interaction layer.

The LLM interaction layer (Fig.~\ref{figs:architecture}-b) is responsible for bridging the gap between natural language ongoing conversation with human operators and underlying data and tools or agents. Architected in a hierarchical fashion, the user interaction, represented as the message history is broken down into subtasks by a top-level Query Processing (QP) Module. These sub-tasks are then “delegated” to lower-level agent(s) and tool(s) based on the content and the context. The sub-tasks are then executed, and their results are recorded back into the message history. 
The QP module will then synthesize the results to respond to the original user query(Fig.~\ref{figs:architecture}-c). 
With this hierarchical architecture,  we have added three modules at the lower level: the \textit{Information Retrieval (IR)} Module, the \textit{Descriptive Analytics (DA)} Module, and the \textit{Predictive Analytics (PA)} Module (Fig.~\ref{figs:architecture}-b). The system can be easily extended by adding more instances at the lower-level without affecting existing lower-level agents or top-level QP module. 
The design provides users with an abstraction layer, enabling them to perform a wide range of on-demand, iterative analytics, including complex multimodal tasks,
without worrying about the complexity.
\subsubsection{Hierarchical Agentic Query Processing Module}
\label{subsec:architecture:agentic_routing}
At the core, this architecture leverages LLM driven agentic query processing and tool calling loops designed to perform specific tasks depending on its role, level, and capabilities given (Fig.~\ref{figs:architecture}-b). Leveraging the reasoning, planning, tool calling, and information synthesizing capabilities, the QP module provides an autonomous recursive interaction between itself and the lower-level agents. This provides the system with a capability 1) to reason about how to coordinate the interaction with the environment and 2) recover from errors if any happened. We have found this capability crucial to bridge the gap between the non-determinism inherent in these LLMs and the determinism that the rest of the system requires.

Further, the “tool calling” capability enables hierarchical design that promotes composability. Instead of a monolithic interaction between the human operator and the underlying environment, the LLM interaction layer is divided into a top-level manager agentic loop for orchestrating delegations and several low-level agentic loops for specific data interactions as shown in boxes in Fig.~\ref{figs:architecture}-b). 
This hierarchical and modular design also enables the \textit{QP} module to respond to complex queries that necessitate interactions with multiple modules. For instance, to determine the number of HPC jobs utilizing GPUs, the \textit{IR} module must first supply the domain information regarding idle power. Subsequently, the \textit{DA} module can filter jobs in the telemetry data that exhibit GPU utilization exceeding this idle power threshold. 
The \textit{QP} module routes the user question to first \textit{IR} and then \textit{DA} agents and then synthesizes the intermediate results from  these modules to finally respond to the user question.

 
  \subsubsection{Information Retrieval Module} 
\begin{figure}[t]
  \centering
  \includegraphics[width=\linewidth]{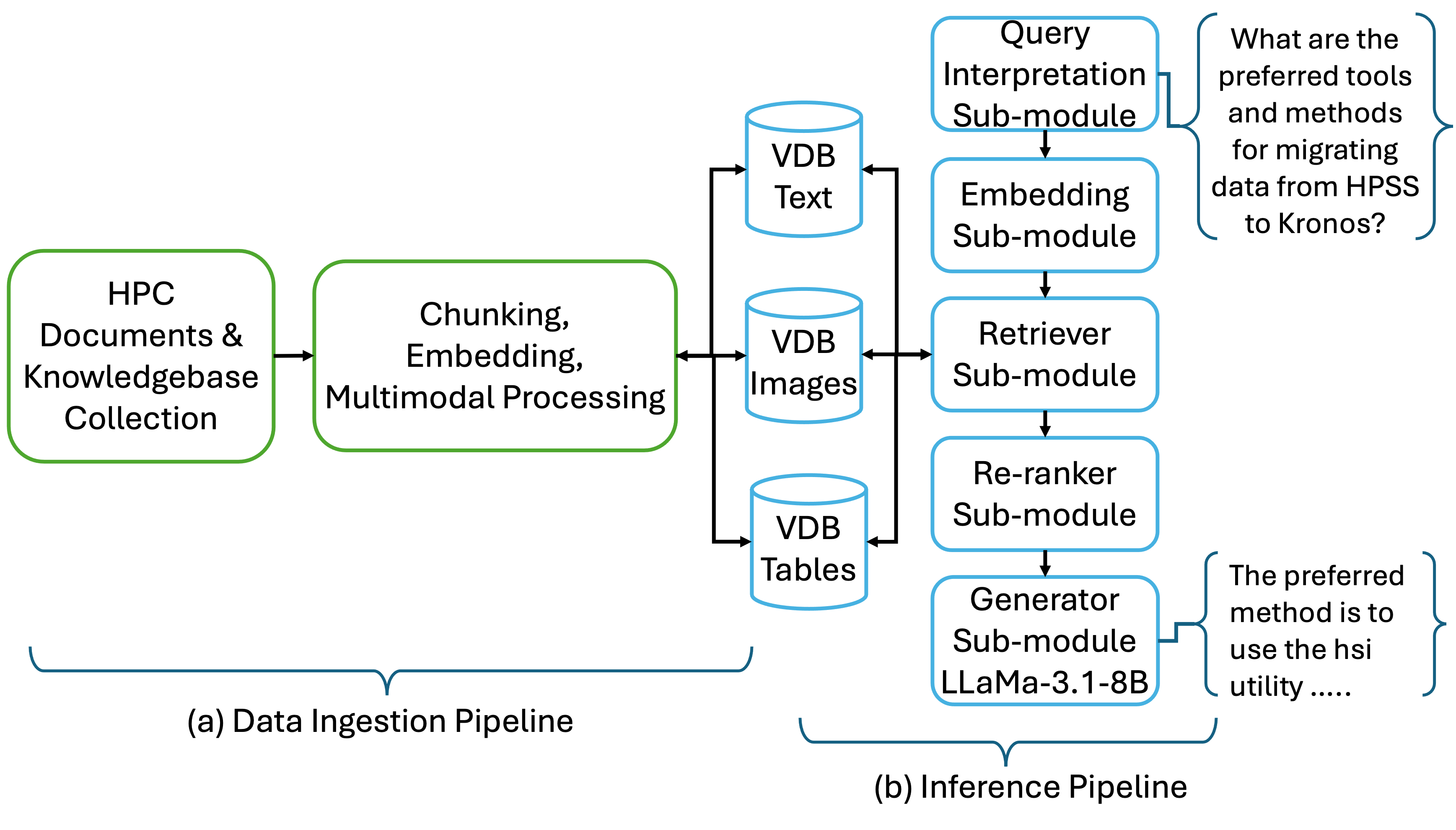}
    \caption{The Document-RAG module: (a) Ingestion pipeline, and (b) Inference pipeline}
  \label{figs: document_module}
\end{figure}

  The Information Retrieval (\textit{IR}) module or agent in the EPIC system is handles HPC-specific queries and retrieves relevant domain information from HPC user manuala, web, academic papers, HPC's auxiliary systems manual. 
  In addition to text responses, the \textit{IR} module also provides structured data in the form of tables and graphical information as images. 
  The \textit{IR} data ingestion pipeline shown in Figure~\ref{figs: document_module} consists of several tasks: chunking, semantic retrieval with hash-based indexing for tables, a reranker for candidate selection, and the \texttt{Llama-3.1-8B} large language model for final response generation. This integrated approach enhances user interaction by returning rich, contextually accurate responses that include both textual and visual data. 

 The \textit{IR} retrieval-augmented generation (RAG) system architecture comprises five integrated sub-modules that transform user queries into contextually relevant responses from HPC documentations. The \textit{IR} module has its own \textit{Query Interpretation (QI)} sub-module that performs decomposition of complex queries into sub-queries and expands wthem with domain-specific terminology. The \textit{IR} module handles content chunking based on semantic coherence and specializes in \textit{multimodal content processing}, extracting tables and images from HPC documents.  
 During the pipeline building phase, the tables from the HPC documentation re processed using \textit{GPT-4o} to generate their descriptive textual representations. These descriptions capture the semantic content and purpose of each table, enabling retrieval based on the information they contain instead of only their surface structure.  For image data, embedded images are extracted using \textit{openai/clip-vit-base-patch32} model, and their captions and additional text is generated with \textit{GPT-4o}. This ensures that diagrams, charts, and other visual elements in the documentation become retrievable during inference phase. 
 The \textit{IR} module leverages ChromaDB as a vector database, employing hybrid retrieval techniques that combine dense vector similarity matching (\textit{mxbai-embed-large-v1}) with sparse retrieval and filtering capabilities. Separate vector databases store table and image embeddings along with their textual descriptors using \textit{Mixed-Bread} \cite{emb2024mxbai} general-purpose light-weight embeddings (512 dimensions). 
 The retrieved content undergoes refinement in the \textit{Re-ranking} sub-module (\texttt{mxbai-rerank-large-v1}), which scores chunks based on semantic relevance, content authority, and information freshness while promoting diversity in results to ensure comprehensive coverage of decomposed queries. This sophisticated inference pipeline ensures that textual content, tables, and embedded images are all accessible through semantic querying. 
 The \textit{Generation} sub-module synthesize coherent responses by integrating retrieved content chunks into an optimized prompt structure. The module's design enables it to seamlessly incorporate information from text, tables, and images into unified responses that directly address user queries about HPC documentation.  

    \subsubsection{Descriptive Analytics (\textit{DA}) Module} 
\begin{figure}[t]
  \centering
  
  \includegraphics[width=\linewidth]{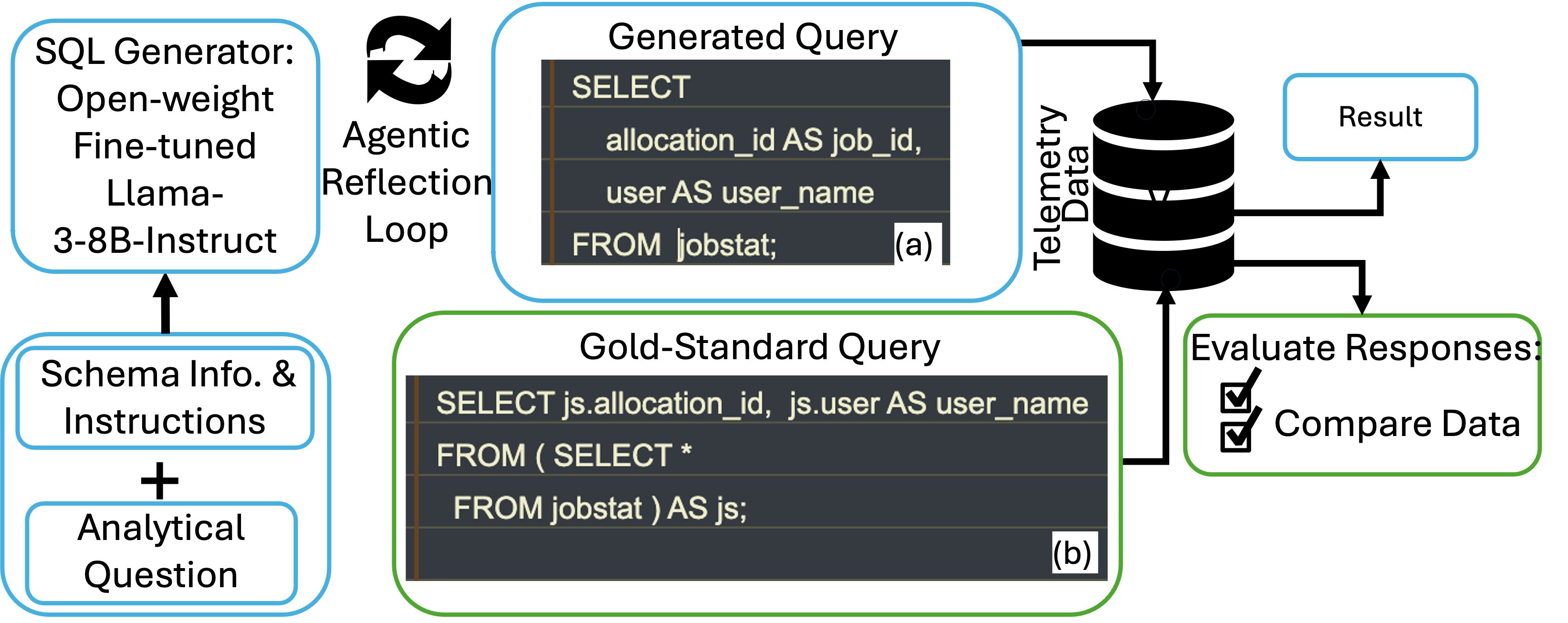}
  \caption{Descriptive Analytics module. The inference pipeline is shown by components in blue outline. The auxiliary pipeline (shown in green) is used in evaluation.}
  \label{fig:telemetry_module_diagram}
\end{figure}

    The \textit{DA} module is designed to facilitate users from diverse backgrounds in performing iterative, exploratory ODA tasks on retrospective telemetry data. 
    Figure~\ref{fig:telemetry_module_diagram} illustrates the various components and the workflow of the DA module pipeline. 

    Whenever a user query is delegated to the DA module, it is concatenated with instructions and schema information. Instructions provide the context to the model, helping it understand the query better and respond effectively. The schema information includes descriptions of telemetry data stored in tables, such as table names, field names, and relationship between tables. The combined input is then processed by the fine-tuned open-weight \texttt{Meta-Llama-3-8B-} \texttt{Instruct} LLM model~\cite{defog_sqlcoder}. The model generates a SQL for the input prompt. The agentic-reflection loop validates the generated SQL for syntax and schema. If the validation fails the model revises the generated SQL, and the process repeats until a valid SQL query is generated or a maximum number of retries is reached. The final SQL query is then executed on the database, providing the response to the analytical query. The user interface, introduced later, offers on-demand interactive graphical visualizations to illustrate the datasets. 
 
    
    \subsubsection{Predictive Analytics (\textit{PA}) Module} 
\begin{figure}[t]
  \centering
  \includegraphics[width=\linewidth]{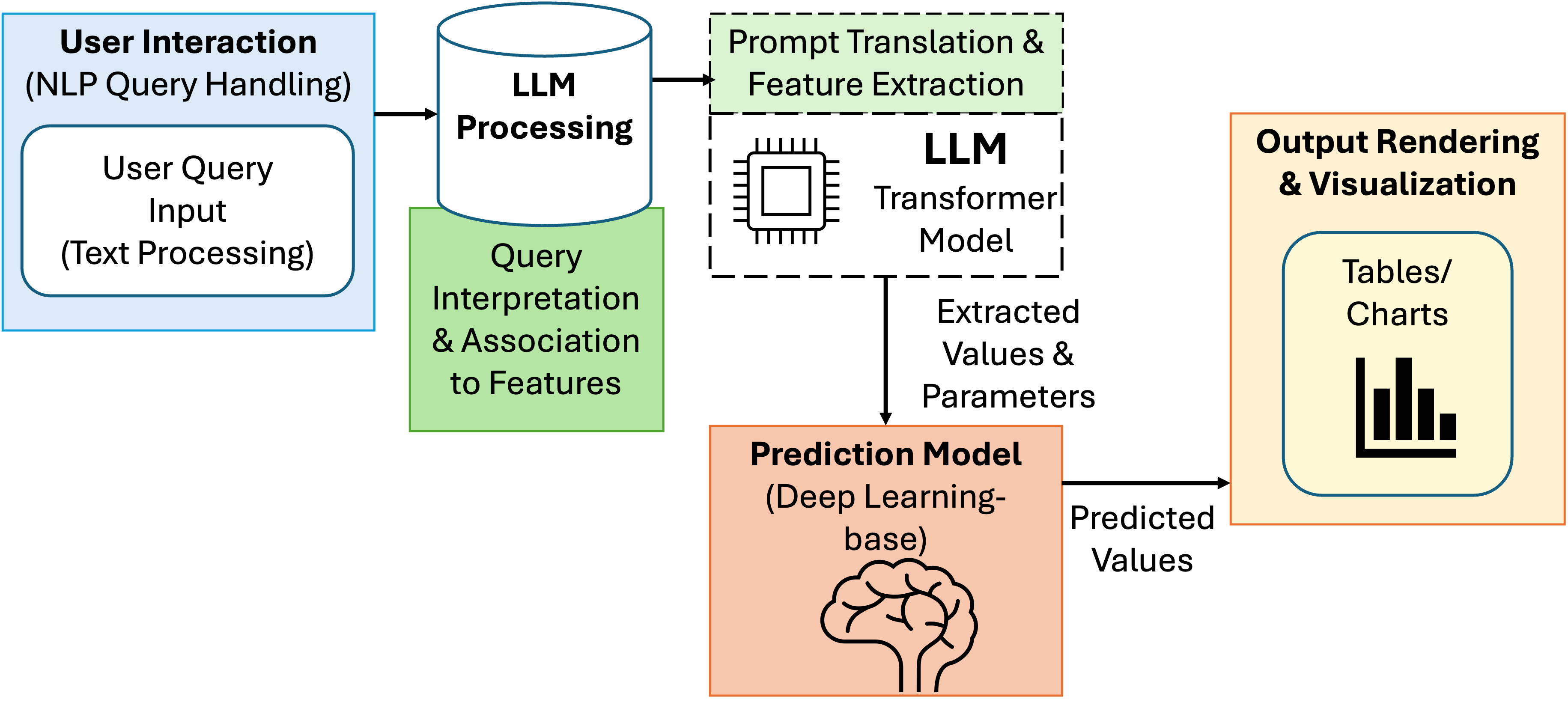}
  \caption{Predictive Analytics Module, showing the interaction between the Large Language Model (LLM), Prediction model, and user interface for generating predictions and visualizations from natural language queries.}
  \label{figs:prediction_diagram}
\end{figure}

     The PA module is designed to handle tasks that estimate HPC job metrics, such as power utilization, energy consumption, and compute node temperature. Figure~\ref{figs:prediction_diagram} illustrates the components and workflow of the prediction module, which consists of two sub-modules.
     
     The first sub-module  is the \textit{Query-Interpretation (QI)}, which interprets user questions directed by the top-level \textit{QP} module to the \textit{PA} module. Similar to the \textit{DA} module, the \textit{QI} agent uses the \texttt{Meta-Llama-3-8B-Instruct} model, fine-tuned for translating natural-language into structured format. 
     Before processing the query, it is combined with information such as science domain names, their descriptions, feature names, and descriptions to assist the model in interpreting the questions correctly. This model then converts the input queries into a structured format, identifying the input and output variables needed for the regression models to estimate the output value. 
     If users do not provide all input features, the model handles this by using a set of the most frequent values for the missing inputs.

     The second sub-module employs regression models, including a feedforward neural network and a regression decision tree, to perform prediction tasks. For instance, the PA module can address queries such as, \textit{"Estimate the total amount of energy consumed by a job that runs on 1,000 compute-nodes, belongs to the computational fluid science domain, and runs for 2 hours."} These regression models are designed to predict one of the 16 features related to power, energy, and temperature, as detailed in Table~\ref{tab:prediction_feature_summary}. The input features include the science domain, walltime, number of nodes, and whether the job utilizes CPU or GPU. A combination of regression models is employed because neural network models provide better accuracy for certain output features, while regression decision tree models outperform neural networks for others. The prediction results are then returned to the top-level model for synthesis and redirection to the user interface for rendering.

\begin{table}[h]
  \caption{Summarized description of the 16 output features predicted by the \texttt{regression} submodule}
  \label{tab:prediction_feature_summary}
  \centering
  \scriptsize
  \begin{tabular}{lp{5cm}}
    
    \hline
 \textbf{Features}                   & \textbf{Description}\\                                            \hline
 \multicolumn{2}{c}{\textbf{(a)~Power}} \\ \hline
Node: Max, Mean, Stddev    & Max/Mean/Stddev of the average node power of each node  \\
         GPU:   Max, Mean, Stddev   & Max/Mean/Stddev of the average GPU power of each node   \\
        CPU Memory Max          & Max of the average CPU memory power of each node        \\ \hline
\multicolumn{2}{c}{\textbf{(b)~Temperature}} \\ \hline
Node: Max, Stddev          & Max/Stddev of the average node temperature of each node \\\hline
 \multicolumn{2}{c}{\textbf{(c)~Energy}} \\ \hline
Total: Node,GPU,CPU Memory & Total Node/GPU/CPU Memory Energy consumed               \\
            Node:   Max, Mean          & Max/Mean of the total node energy of each node          \\
            GPU:   Max, Mean           & Max/Mean of the total GPU energy of each node  \\
\hline
\end{tabular}
\end{table}

  \subsubsection{Multi-modal Rendering User Interface}
    
    Users interact with the EPIC system via a chat bot style web UI. The user can prompt the model and get text responses, tables, and/or plots in response. Using the agentic routing, the UI allows the user to interact with the \textit{IR}, \textit{DA}, and \textit{PA} modules as needed based on their query.
    Unlike conventional chat bots, we chose to make the web UI output plots as dynamic widgets. Users can interact with the widgets to change the plot type and data columns rendered, e.g. to switch between line and bar charts, or to select a specific portion of the data which is relevant. This flexiblity allows users to easily customize the plot rendering to fit their specific use case.
     
    We wanted to avoid the AI agent being a "black box", and to allow users to inspect the agent's reasoning process. The UI displays intermediate steps and tool calls the agent makes in expandable sections in the chat. In addition, if the user prompt requires querying the \textit{DA} module to answer, the generated SQL can be shown in a side window along with the tables and plots, at the user's discretion.
  
\subsection{Evaluation Pipeline}
    Evaluating the EPIC system is essential for assessing its quality and fostering confidence in its wide-ranging, multifaceted HPC operational analytics capabilities. We present an extensive analysis of the \textit{IR}, \textit{DA}, and \textit{PA} modules, along with an evaluation of the top-level Query Processing Module. This includes an end-to-end assessment to demonstrate combined performance on hybrid questions that require multiple tool calls. We detail the extensive evaluation results in Sec.~\ref{sec:evaluation}.  
    
    \textit{High-level Reasoning, Routing, and Synthesis Evaluation~} 
    We evaluate the top-level hierarchical routing described in Sec.~\ref{subsec:architecture:framework}, a critical element for composability and extensibility. 
    Our assessment focuses on whether the model can accurately "route(delegate)" user requests to the correct low-level agents and synthesize the results. We report metrics including routing accuracy, synthesis accuracy, and total token cost. 

    \textit{Information Retrieval Module Evaluation: }
    Evaluation of the multi-modal Retrieval Augmented Generation (RAG) system ensures robust and contextually relevant retrieval and generation of HPC knowledge-base and documentations; please refer to Section 5.1 for details. 
    %
     
     \textit{Descriptive Analytics Module Evaluation:~} 
     We evaluate the \textit{DA} agent by comparing the datasets extracted from generated SQL queries against the datasets generated by gold-standard queries. 
     This approach is selected because multiple correct SQL statements can produce the same dataset, as illustrated by queries (a) and (b) in Figure~\ref{fig:telemetry_module_diagram}.
    
    However, evaluating datasets rather than SQL statements presents several challenges. Column names may differ between datasets even if the data is identical. Additionally, the order of data rows can vary, which is important for tasks like retrieving the top k rows. Furthermore, datasets may have matching values but differ in row-to-row mapping. 
    To address these challenges, we mark a response as correct only when the data in the columns matches the data generated by the gold-standard queries, regardless of column names, and when the row order and mapping are accurate.
     \textit{Predictive Analytics Module Evaluation:~}
     The \texttt{PA} module comprises two main components: the \texttt{query interpretation} sub-module and the \texttt{regression} module. For query interpretation, we use the accuracy metric to assess how well the model converts user input into structured JSON. Specifically, we evaluate the accuracy of each of the four input features and the output feature, which can be any one of the 16 features (refer to Figure 1). For evaluating the prediction regression sub-module, we employ three error metrics to assess the model's performance: Mean Absolute Error (MAE), Root Mean Square Error (RMSE), and Mean Absolute Percentage Error (MAPE).

\section{EPIC Software \& Data}
\label{sec:data}
    In this section we describe the EPIC platform software and datasets that we used for evaluating the various modules.
\begin{table}[t]
    \caption{EPIC: Software, Models and Data}
    \label{tab:models_data}
    \footnotesize
    \addtolength{\tabcolsep}{-0.2em}
    \begin{tabular}{ll}
        \toprule
        
        \textbf{Dataset} & \textbf{Description}\\
        \midrule
        {(i)~\textit{IR} Module}          & Expert vetted, gold-standard set of 1000 Q\&A from \\
        & the OLCF documents, manuals, Wikis, and articles.\\
        {(ii)~\textit{DA} Module}       & About 500 pairs of english language and SQL queries. \\
        &Database schema of Frontier telemetry data. \\
        & Generated telemetry data for evaluating the \textit{DA} module.\\
        {(iii)~\textit{PA} Module}      & (a) Vetted 100 prediction Q\&A  (in 'json' format). \\  
        & (b)Regression models evaluated on Frontier data,\\ 
        & 1-year jobs log containing about 0.5 million jobs. \\ 
        {(iv)~Top-level}  & 
        Generated $181$ data points for the agentic routing \\
        ~\textit{QP} Module & and synthesis evaluation. \\
        \bottomrule        
                                            
    \end{tabular}
\end{table}
    The top-level \textit{QP} module consist of agentic loop to call lower-level agents. 
    Each of the modules are implemented using LangChain v0.3~\cite{langchain}.  
    The \textit{IR} module stores documents in ChromaDB v0.6~\cite{chromadb}, the \textit{DA} module's telemetry database uses the library based (DuckDB v1.1~\cite{duckdb}), and the regression models are trained using Pytorch v2.6~\cite{pytorch}.  All the components are integrated together in a Chainlit v2.2~\cite{chainlit} based chat UI.  
    Data for each access point is ingested into the system via out-of-band pipelines, separate from the hierarchical LLM interaction layer, (see Figure~\ref{figs:architecture}).

    Due to the varied natural language capabilities, each component leverages different underlying LLMs. 
    This resulted in a several Llama~3 based fine-tuned models \cite{grattafiori2024llama3herdmodels} for low-level components and OpenAI models \cite{openai2024gpt4ocard} for the high-level reasoning, routing, and synthesis.

    The EPIC design allow each component can be imported separately from an out-of-band evaluation pipeline, enabling batch end-to-end tests without the WebUI. With the data already ingested, the code-paths for each component in the LLM interaction layer has been evaluated as if deployed in a real environment. 

    Table~\ref{tab:models_data}, summarizes the dataset prepared for evaluating all components of the EPIC platform. 
    For evaluating \textit{IR} module, we curated a \textit{gold-standard} set of Q\&A pairs, written by subject-matter experts to simulate real HPC center queries. The Q\&A pairs include textual data, as well as tables and images from documentations. 
     
    The \textit{DA} evaluation dataset contains about 500 natural language questions and SQL response pairs. We also maintain a generated telemetry dataset schema for performing analytics on energy, power, temperature and their corresponding scheduler logs. 
    The schema definition data, including table names, column descriptions, data types, and table relations is important part of the input prompt alongside natural language questions.

    The~\textit{QI} sub-module of the Predictive Analytics module evaluation dataset includes 100 prediction Q\&A pairs related to power, temperature, and energy consumption of supercomputer jobs. The ground-truth responses are structured in JSON format. For the regression sub-module, we used 80\% of one year's worth of $0.5$ million Frontier job-level data for training and 20\% for evaluation.
\section{
    Empirical Evaluation }
    \label{sec:evaluation}
        Using the application software and data, we conducted extensive quantitative evaluations of each module, including top-level end-to-end evaluations for synthesis and routing. The LLM models across different modules have varying inputs and outputs, leading to distinct metric criteria. These criteria validate various aspects of the EPIC modules.
        

\begin{table}[]
    \caption{Evaluation of the document module via Lexical, Semantic, Hallucination metrics}
    \label{tab:eval}
    \scriptsize
    \centering
    \addtolength{\tabcolsep}{2em}
    \begin{tabular}{l l c}
        \hline
        \textbf{Category} & \textbf{Metric} & \textbf{Value} \\
        \hline
        \multicolumn{3}{c}{\textbf{Lexical} (higher better)} \\
        \hline
        \quad Rouge1 & F1 & \color{red}{0.58} \\
        & Precision & \color{red}{0.55} \\
        & Recall & \color{red}{0.62} \\
        \cline{2-3}
        \quad RougeL & F1 & \color{red}{0.50} \\
        & Precision & \color{red}{0.47} \\
        & Recall & \color{red}{0.53} \\
        \hline
        \multicolumn{3}{c}{\textbf{Semantic} (higher better; except BARTScore)} \\
        \hline
        \quad BERT & F1 & \color{red}{0.66} \\
        & Precision & \color{red}{0.72} \\
        & Recall & \color{red}{0.58} \\
        \cline{2-3}
        & BART & \color{red}{-2.39} \\
        \hline
        \multicolumn{3}{c}{\textbf{Hallucination} (lower better)} \\
        \hline
        & SelfCheckGPT-NLI & \color{red}{0.33} \\
        \hline
    \end{tabular}
    \label{tab:doceval}
\end{table}

        \subsection{Information Retrieval Module} Our Information Retrieval module is a Retrieval Augmented Generation (RAG) system for HPC-centric question-answering. The responses from this multimodal module is critical for user-queries on HPC documentation and knowledge.
    
        ~\textit{Lexical and Semantic evaluation}: We employ standard Natural Language Processing (NLP) metrics to evaluate the answers generated by the \textit{IR} module.
        To capture both lexical and semantic essence, we use ROUGE metrics \cite{lin-hovy-2003-automatic} for lexical and BERTScore \cite{DBLP:conf/iclr/ZhangKWWA20}; BARTScore \cite{10.5555/3540261.3542349} for semantic similarity with ground-truth data. 
        The objective is to assess how well our generated answers reflect the actual answers for the given queries at both lexical and semantic levels.
        For lexical evaluation (Table \ref{tab:doceval}), we use ROUGE-1 and ROUGE-L, as fluency and coherence (ROUGE-2) are captured in our human and LLM-as-judge evaluations.
        The lexical evaluation (via ROUGE scores; ROUGE1 and ROUGE-L) reveals that the system has a performance-affinity towards generating concise answers withhigh precision but low recall, effectively avoiding noise but sometimes misses relevant content.
        
        However, lexical metrics do not capture semantic similarity (paraphrases, synonyms) and can penalize valid answers if phrasing differs, and can be gamed by long answers with keyword stuffing. 
        Hence, we perform semantic-level evaluation of our system-generated answers against gold-standard. 
        BERTScore is a semantic-level Natural Language Generation (NLG) evaluation metric based on token-level Precision, Recall, and F1, utilizing contextual embeddings from the BERT language model \cite{devlin-etal-2019-bert}. It compares predicted and reference texts using contextual embeddings, capturing meaning beyond exact word overlap. This make BERTScore well-suited for evaluating paraphrased or abstractive QA answers. From Table \ref{tab:doceval}, BERTScore-F1 is 0.62 which is considered to be \textit{good} with some paraphrasing or minor errors. 
        However, like the lexical metrics we observe high precision and low recall situation. This signifies that our generated answers are concise, and hence may not be adequate enough in some cases.
 
        We also employ BARTScore, a semantic evaluation metric inspired by text-generation model BART \cite{lewis-etal-2020-bart}, commonly used for longform QA evaluation. 
        BART score measures the 
        likelihood of reference text given the generated text providing a more semantically aware than ROUGE or BLEU. However, BARTScore values are negative log-likelihoods, so less negative indicate better performance. 
        From {Table \ref{tab:doceval}, BARTScore-F1 is -4.12 which, while not ideal, reflect the concise nature of generated answers. 
        This score suggest, generated answers is likely relevant and partially correct, but may be missing details, consistent with the lexical evaluation.}
        
        One critical issue with Generative AI models is \textit{Hallucination}, where generated answers sound plausible but contain incorrect information. 
        Lexical and semantic NLG metrics do not verify factual correctness, only linguistic similarity. 
        To address this, we combine these metrics with faithfulness checks, such as entailment models or human judgment. 
        We use the SelfCheckGPT-NLI \cite{manakul-etal-2023-selfcheckgpt}, an automated Natural Language Inference (NLI)-based metric for hallucination evaluation.
        SelfCheckGPT-NLI detects hallucinations by measuring contradictions between generated answers and multiple sampled outputs.
        A higher contradiction score indicates lower factual consistency. 
        A SelfCheckGPT-NLI score of 0.45 signifies a moderate-low likelihood of hallucination in the QA answer.
        
        \textit{Multimodal Information Retrieval: }We evaluate the multimodal information retrieval capability of the \textit{IR} module in Table \ref{tab:table-image-retrieval}, using metrics like Top-k (k=1) Accuracy and Mean Average Precision (MAP).
        These metrics asses whethercorrect images and tables are retrieved. 
        We penalize responses that generate only text-based answers for multimodal questions. 
        The Top-1 accuracy metric is `strict/hard match' and a score of 0.68 for image and 0.65 for table, indicating good performance. 
        MAP scores (lenient than Top-1) are based on retrieval @ 2 (see Table \ref{tab:table-image-retrieval}) indicates precise retrieval of relevant construct (table/image) in response to user queries.
        
        \textit{Human Evaluation and LLM-as-judge}: Automated NLP metrics are still not adequate to cover the full spectrum of evaluation. 
        Hence, we employ \textit{human evaluation} using criteria such as \textit{Correctness, Completeness, Fluency, Relevance} on a Likert scale of 1-5. Table \ref{tab:human-llm-eval}.
        An HPC domain expert rated the generated answers, finding them highly coherent and relevant, with strong-alignment to input queries. Relative lower factual accuracy indicate occasional inconsistencies, Completeness core indicate reasonable coverage, with room for improvement.  
        Additionally, we used state-of-the-art Large Language Model GPT-4o to evaluate the generated answers with the same criteria as the domain experts on a randomly sampled 50 QA pairs (Table \ref{tab:human-llm-eval}).
        The results show high high relevance and coherence metrics value but relatively lower on factual accuracy and completeness. 
        
        \textbf{Finding 1:} \textit{The multimodal \textit{IR} module performs well with shorter Q\&A, generating relevant and correct responses; however  there is room for improvement in handling long-form Q\&A. The results underscore the importance of extensive evaluation.} 

        \begin{table}
  \caption{Table and Image Retrieval Performance}
  \label{tab:table-image-retrieval}
  \footnotesize
  \centering
  \begin{tabular}{ccccc}
    
    \toprule
      \textbf{Modality} & \textbf{Top-1 Accuracy} & \textbf{MAP}   \\
    \midrule
      Image & 0.68 & 0.72  \\

    \midrule

Table & \color{red}{0.66} & \color{red}{0.68}  \\

  \bottomrule
\end{tabular}
\end{table}

    \begin{table}
  \caption{Human and LLM-as-judge Evaluation}
  \label{tab:human-llm-eval}
  \footnotesize
  \addtolength{\tabcolsep}{-0.4em}
  \centering
  \begin{tabular}{ccccc}
    
    \toprule
      \textbf{Participant} & \textbf{Relevance} & \textbf{Coherence} &  \textbf{Fac. Accuracy} & \textbf{Completeness}\\
     
    \midrule
      SME & 3.92 & 4.15 & 3.65 & 3.76 \\
     

    GPT-4o & 4.19 & 4.18 & 3.74 & 3.41 \\

  \bottomrule
\end{tabular}
\end{table}
    
    \subsection{Descriptive Analytics}
        The \textit{DA} module is primarily based on the conversion of natural language into SQL for operational analytics tasks for insights on the past datasets. 
        \begin{table}
      \caption{Evaluation of the telemetry results based on the complexity by SQL functions.}
      \label{tab:telemetry_results_key_fun}
    \scriptsize
    \addtolength{\tabcolsep}{-0.4em}
    \begin{tabular}{llcc}
    \toprule

         & \textbf{{SQL Functions \& Keywords}} & \textbf{All} & \textbf{Correct} \\
            \midrule
        0 & AVERAGE, MAX, JOIN, GROUP BY & 11 & 8 \\
        1 & CASE WHEN, COUNT, GROUP BY, NEW COLUMN & 11 & 8 \\
        2 & COUNT, AVERAGE, EXTRACT, JOIN, GROUP BY & 11 & 10 \\
        3 & COUNT, DISTINCT & 7 & 6 \\
        4 & COUNT, FRACTION, WHERE & 11 & 11 \\
        5 & COUNT, FRACTION, WHERE, JOIN, CASE WHEN, GROUP BY & 11 & 9 \\
        6 & COUNT, JOIN, GROUP BY & 31 & 27 \\
        7 & COUNT, JOIN, WHERE, GROUP BY & 24 & 22 \\
        8 & COUNT, JOIN, WHERE, GROUP BY, ORDER BY & 22 & 20 \\
        9 & COUNT, WHERE & 11 & 11 \\
        10 & DATE\_TRUNC, AVERAGE, JOIN, GROUP BY & 11 & 11 \\
        11 & DATE\_TRUNC, AVERAGE, JOIN, GROUP BY, ORDER BY & 22 & 17 \\
        12 & DATE\_TRUNC, AVERAGE, MAX, WHERE, GROUP BY, ORDER BY & 11 & 11 \\
        13 & DATE\_TRUNC, COUNT, GROUP BY & 11 & 11 \\
        14 & DATE\_TRUNC, COUNT, GROUP BY, ORDER BY & 22 & 21 \\
        15 & DATE\_TRUNC, COUNT, JOIN, GROUP BY & 11 & 11 \\
        16 & DATE\_TRUNC, COUNT, JOIN, WHERE, GROUP BY, ORDER BY & 11 & 11 \\
        17 & DATE\_TRUNC, COUNT, WHERE, GROUP BY & 11 & 11 \\
        18 & DATE\_TRUNC, MAX, GROUP BY & 22 & 19 \\
        19 & DATE\_TRUNC, SUM, JOIN, GROUP BY & 11 & 10 \\
        20 & DATE\_TRUNC, SUM, JOIN, GROUP BY, ORDER BY & 11 & 11 \\
        21 & DATE\_TRUNC, WHERE, GROUP BY & 11 & 10 \\
        22 & DISTINCT & 9 & 8 \\
        23 & EXTRACT, AVERAGE, WHERE, DATE\_TRUNC, GROUP BY & 11 & 1 \\
        24 & JOIN, GROUP BY, ORDER BY & 22 & 22 \\
        25 & SUM, FRACTION, JOIN, GROUP BY & 11 & 0 \\
        26 & SUM, GROUP BY & 33 & 29 \\
        27 & SUM, JOIN, GROUP BY & 22 & 20 \\
        28 & SUM, WHERE, GROUP BY & 22 & 22 \\
        29 & WHERE, ORDER BY, LIMIT & 8 & 8 \\
        \bottomrule
    \end{tabular}
    \end{table}
    
    

    
        For evaluation, we compare 453 generated SQL statements with the ground-truth counterparts, grouping them by keywords and functions (shown in~Table~\ref{tab:telemetry_results_key_fun}). 
        Using the evaluation method described in Section~\ref{sec:design}.2, the model generated correct SQL for 396 data points.  
        Grouping help us ascertain the quality of the model based on complexity of SQL. Complex of SQL depends on number of \textit{keywords} and type of \textit{keywords}. 
        Table~\ref{tab:telemetry_results_key_fun} lists 30 different SQL patterns, showing the number of questions per pattern and the count of correctly converted natural language queries. Some patterns exhibit high matching scores, while others, particularly those in rows 23 and 25, have significantly lower success rates.  
        
        We further analyze these \textit{SQL Functions and Keywords}, which are compound patterns of multiple keywords, by decomposing them into atomic patterns. This involves extracting and analyzing individual keywords.

\begin{figure}[t]
          \centering
          \includegraphics[width=\linewidth]{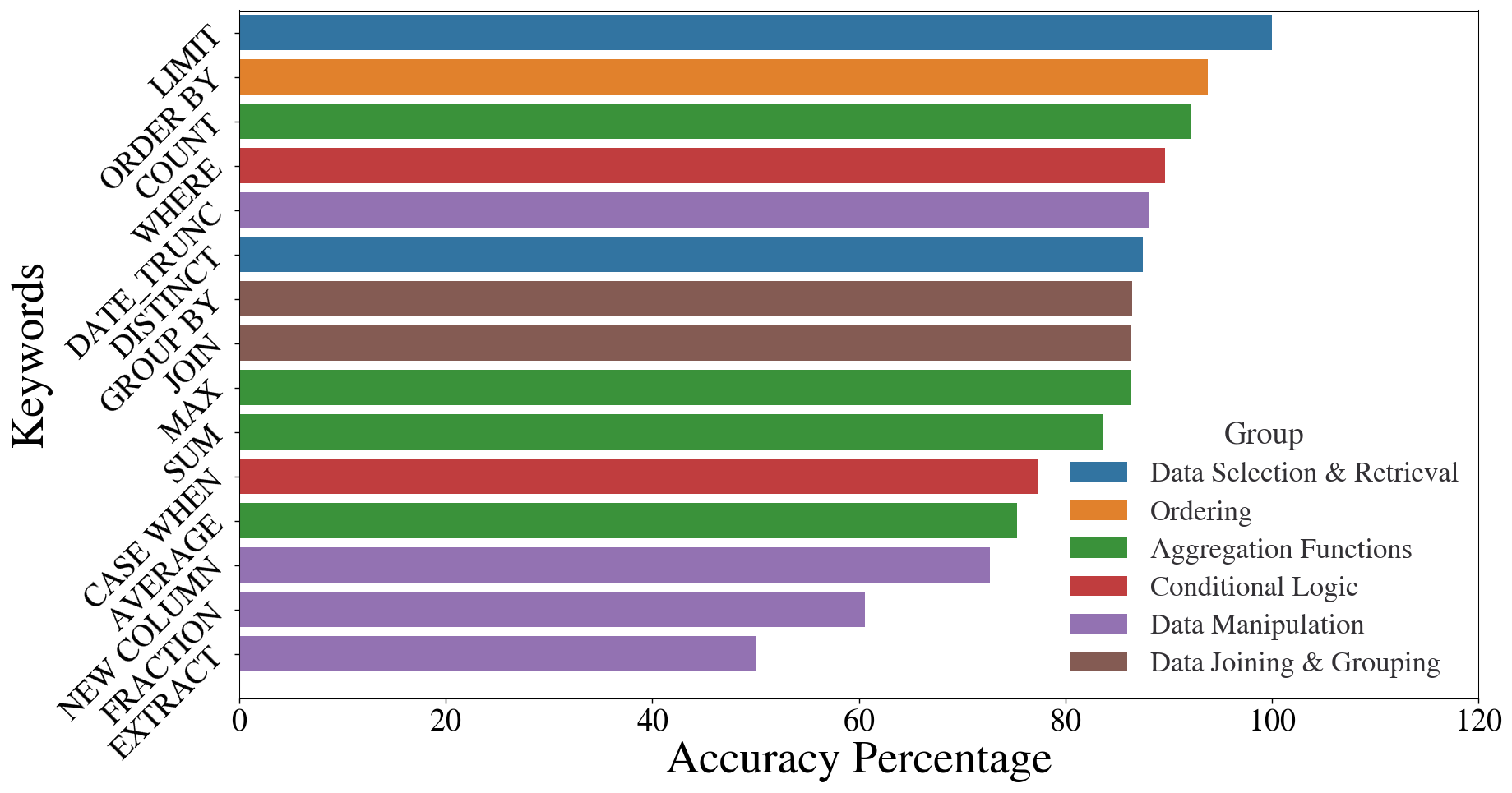}
          \caption{SQL keywords wise performance of the telemtery module. 
          The keywords are classified into six groups based on the type of SQL keywords. }
          \label{figs:telemetry_group_accuracy}
        \end{figure}

        The plot in Figure~\ref{figs:telemetry_group_accuracy} shows the model performance based on the atomic \textit{Keywords} vs. accuracy. Atomic patterns, which involve single keywords reveal that functions, such as \textit{LIMIT} and \textit{ORDER BY}, have the highest accuracy, while \textit{EXTRACT} \& \textit{FRACTION} have the lowest accuracy percentage.  \textit{EXTRACT} is a SQL function to extract used to extract absolute time component, such as the hour of the day ($0$-$23$), or week of the year ($0$-$53$), independent of the year. 
        We then group these keywords into six categories based on the type of SQL function types: \textit{selection \& retrieval}, \textit{aggregation functions}, \textit{data manipulation}, \textit{conditional logic}, \textit{joining \& grouping}, and \textit{ordering}.
        The plot in Figure~\ref{figs:telemetry_group_accuracy} also illustrates the performance of the model'sperformance across these six categories.We observe that the data manipulation functions are relatively complex and more challenging to generate. 
 
\begin{figure}[t]
          \centering
          \includegraphics[width=\linewidth]{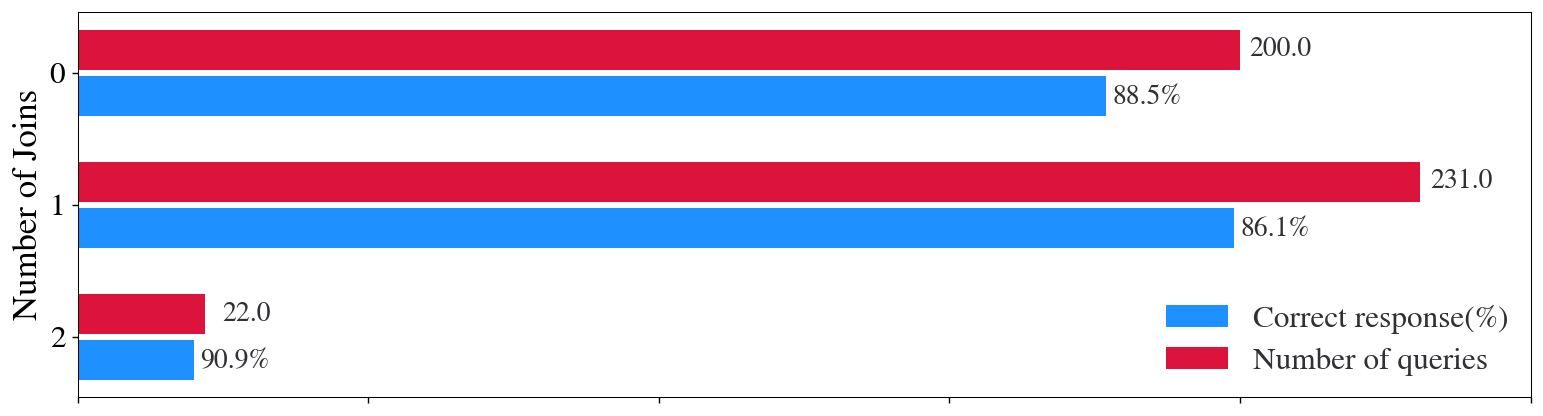}
          \caption{Model performance based on the number of data source involved.}
          \label{figs:telemetry_join_accuracy}
        \end{figure}

        The number of table joins or combined dataset sources is a crucial aspect of analytical questions. Increasing the number of joins can enhance query complexity and may affect the model performance. Figure~\ref{figs:telemetry_join_accuracy} illustrates the model's accuracy based on the number of joins. The red bar indicates number of questions for each 'join' category, while blue bar shows the percentage accuracy for the respective number of joins. 
        We observe that model performs well across all the three categories, maintainin at least 85\% accuracy. This is significant, as the HPC ODA tasks often require merging data from multiple tables, and the model strong performance in handling joins is encouraging.  
        
\begin{figure}[t]
          \centering
          \includegraphics[width=\linewidth]{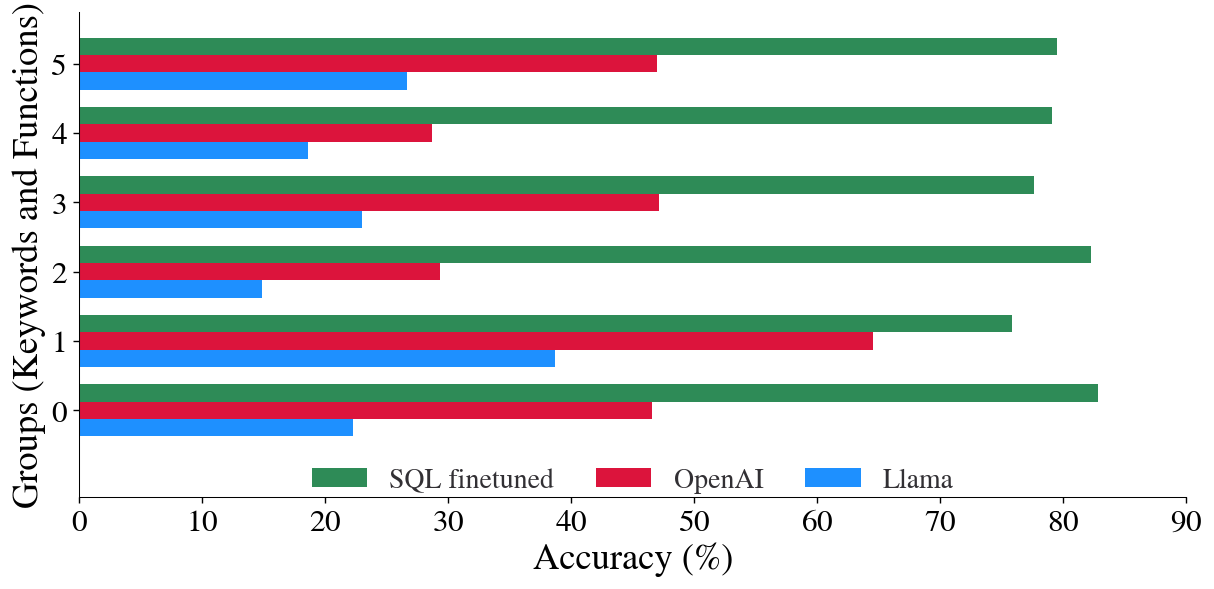}
          \caption{Performance of three models. (0: Aggregation functions, 1: Conditional logic,
          2: Data manipulations, 3: Joining and Grouping, 4: Ordering, and 5: Selection and retrieval)}
          \label{fig:telemetry_model_comparision}
        \end{figure}

        Generating SQL is not a typical text-to-text generation problem for language models. We compare the general purpose state of the art propriety \textit{OpenAI-4o} model with open-weight \textit{Llama-3 model} model fine-tuned for SQL. The  \textit{OpenAI-4o} model is significantly larger than 8-billion parameter \textit{Llama-3} model we employed. Figure~\ref{fig:telemetry_model_comparision} compares the 3 models: base \textit{Llama-3-8b} model, \textit{OpenAI-4o} model and \textit{Llama-3} 8b model fine-tuned for SQL datasets. While \textit{OpenAI-4o} model outperforms the base \textit{Llama-3} model, the fine-tuned \textit{Llama-3} model exhibits significantly better accuracy than the \textit{OpenAI-4o} model for SQL tasks. 
        
        \textbf{Finding 2: } \textit{The accuracy of the analytical model may depend on the complexity of the SQL being generated. However, we observed that the number of joins did not significantly affect model performance. Users should be cognizant of these nuances when performing analytics with these tools.}
        
        \textbf{Finding 3: } \textit{We demonstrate that for specialized tasks, such as analytics, smaller, fine-tuned models can outperform larger, general-purpose models.}
        
    \subsection{Predictive Analytics}
        In this section, we evaluate the performance of the predictive analytics module, divided into two parts based on the two sub-modules. The first part evaluates the parsing of user questions, and their translation into a structured output by LLM. The second part is evaluates the regression model using various error metrics.
        \begin{figure}[t]
          \centering
          \includegraphics[width=0.7\linewidth]{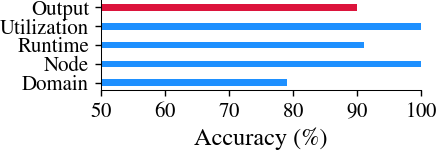}
          \caption{Performance of the LLM model for translating user questions into the input and output feature for the prediction model }
          \label{figs:prediction_llm_eval}
        \end{figure}    
        
        For the evaluation of the natural language parsing, we observe the LLM accuracy in handling four input features and the output feature, as shown in Figure \ref{figs:prediction_llm_eval}.  The model achieves $100\%$ accuracy for two  input features (node count, utilization type) and above $79\%$ accuracy for the remaining two (domain, time). In case of choosing domains, model chooses some irrelevant domain along with the correct domain, which we consider inaccurate, and thus the accuracy drops. Similarly, for the runtime of the job, the model sometimes struggle converting minutes to seconds. In identifying the output feature, the model acheives $90\%$ accuracy. The only errors that have been observed is that in some of the power prediction questions, the model suggests energy as the output feature.

\begin{table}
  \caption{Evaluation of \textit{PA} \texttt{regression} submodule for Test Set}
  \label{tab:prediction_regression}
  \scriptsize
  \centering
  \begin{tabular}{ccccc}
    
    \toprule
      \textbf{Output Features}&\textbf{MAE} & \textbf{RMSE} & \textbf{MAPE}  \\
    \midrule
      GPU Power Max        & 133.60   & 67.23    & 0.35 \\
GPU Power Mean       & 177.01   & 66.38    & 0.34 \\
CPU Memory Power Max & 8.33     & 4.84     & 0.06 \\
Node Power Mean      & 7263.45  & 368.99   & 0.29 \\
Node Temperature Max & 8.96     & 4.58     & 0.08 \\
Total Node Energy        & 9.93E+10 & 6.43E+08 & 0.33 \\
Node Energy Mean     & 2.78E+08 & 1.54E+06 & 0.31 \\
CPU Memory Energy    & 5.08E+07 & 1.32E+06 & 0.12\\

  \bottomrule
\end{tabular}
\end{table}

        \begin{figure}[h]
          \centering
          \includegraphics[width=\linewidth]{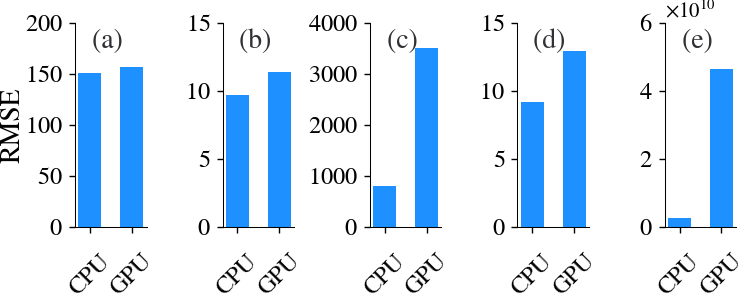}
          \caption{Performance (RMSE) of the \textit{regression}  submodule by utilization type for different output features: (a) GPU Power Max, (b) CPU Memory Power Max, (c) Node Power Mean, (d) Node Temperature Max, (e) Total Node Energy.}
          \label{figs:prediction_rmse_utilization}
        \end{figure}
        
\begin{figure}[t]
          \centering
          \includegraphics[width=\linewidth]{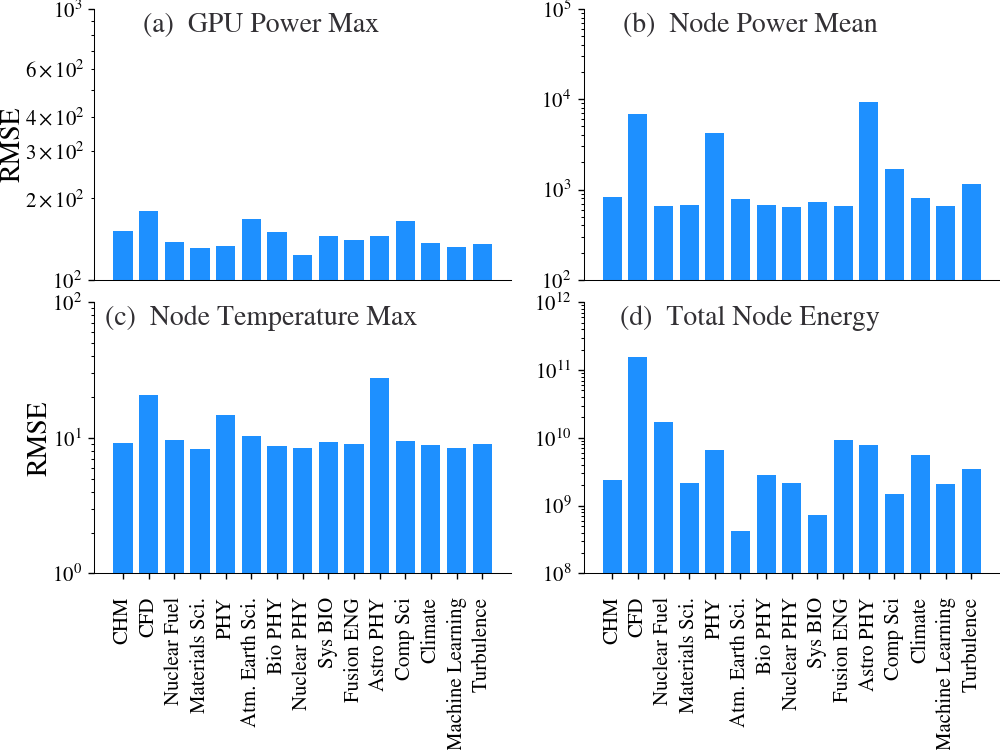}
          \caption{~Performance (RMSE) of the  \textit{regression} submodule by science domains for different output features. }
          \label{figs:prediction_rmse_domain}
        \end{figure}

        For the \textit{regression} submodule evaluation, we report all 
        error metrics for our test data for individual output features in Table \ref{tab:prediction_regression}. We observe that the temperature feature has the lowest error of $8\%$. For the power and energy features, the model shows $65\%$ to $94\%$ accuracy depending on the output feature variable. We also observe how the model performs corresponding to the input feature values. For the jobs utilizing GPUs, it show higher RMSE compared to the jobs running on CPUs given in Figure \ref{figs:prediction_rmse_utilization}. Next, we observe how the model performs for each of the science domains given in Figure \ref{figs:prediction_rmse_domain}. We observe that for the GPU Power or Temperature feature the model performs similarly for all of the domains. In case of the Node Power, and Node Energy we see the performance drop for a few domains namely, Fluid Dynamics (CFD), and Physics for Node Power. In addition, Nuclear Fuel, Fusion Engineering, Astro Physics have higher errors compared to the rest of the domains while predicting Total Node Energy.   \\

        \textbf{Finding 4: }\textit{
        Again, the Llama model demonstrates high performance for the "specific" task of generating structured 'json' datasets.}

        \textbf{Finding 5: }\textit{Predictive models, like the regression model we have demonstrated, can be a modality in operational data representing a summary of past data and can be effectively synthesized into the analysis thanks to the EPIC's modular capabilities.}

    \subsection{Routing, Reasoning, and Synthesis}
    \label{subsec:eval:routing}
        
        In this section, we evaluate the routing, reasoning, and synthesis capability of the high-level agentic tool calling loop.  For this, we have developed an evaluation loop that performs end-to-end tests with test examples that aims for varying degrees of routing and synthesis required by the examples framing it as a classification problem.
        
        In simple cases, the examples would require a single instance of delegation representing RAG – information retrieval, SQL – descriptive analytics, or PRD – predictive analytics as classes.  More complicated cases would demand reasoning and sequencing of multiple capabilities represented as additional four classes: RAG+SQL, RAG+PRD, SQL+PRD, and RAG+SQL+PRD). Additionally, we consider a class which none of the lower-level modules are involved (i.e., greetings in a conversation). The evaluation loop scores both the accuracy of the call and the synthesized result. We employed total 181 examples in which simple case examples are coming from other per-capability evaluations (60 examples in total) and the complex ones were dedicated just for this evaluation.

\begin{figure}[t]
          \centering
          \includegraphics[width=\linewidth]{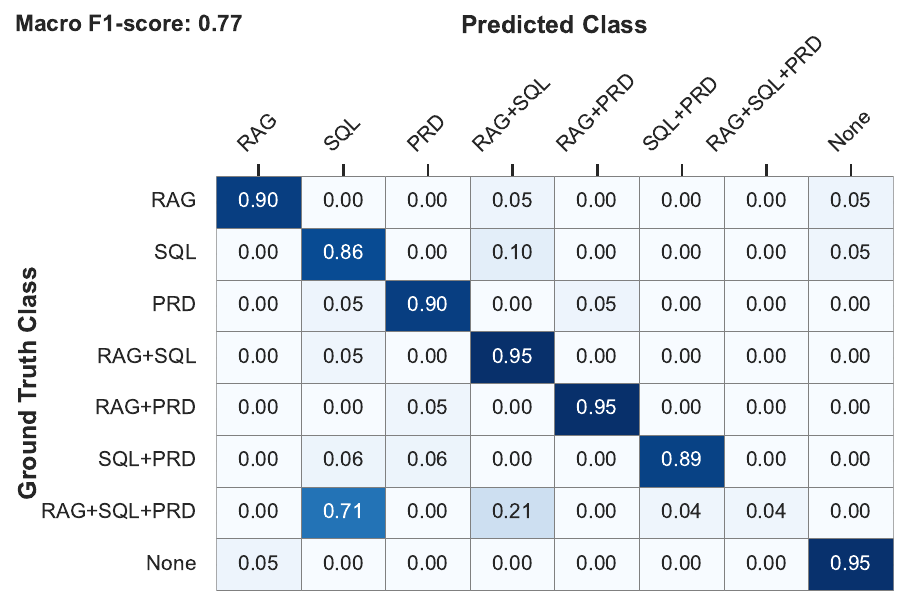}
          \caption{Confusion matrix of the routing capability reasoned by the top-level model yielding an F1 score of 0.77}
          \label{figs:routing_eval}
        \end{figure}

\begin{figure}[t]
          \centering
          \includegraphics[width=\linewidth]{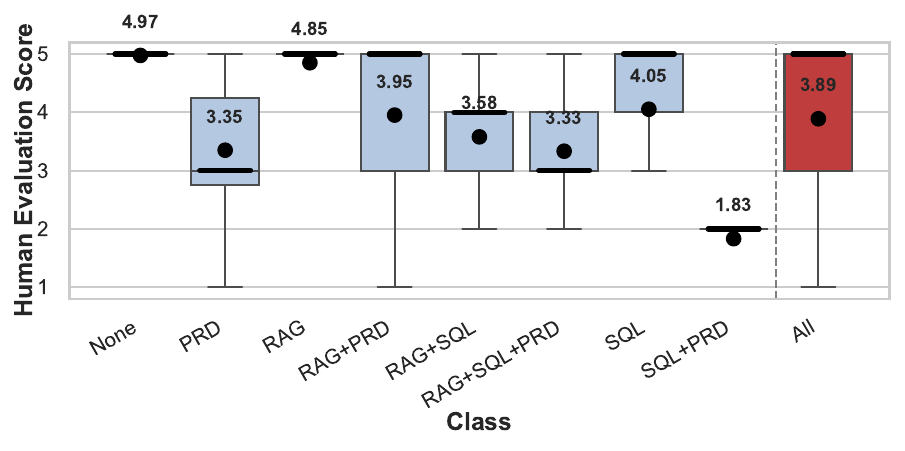}
          \caption{End-to-end human evaluation: performed on 181 high-level questions, reporting mean Likert scale per class (dot) and their distribution}
          \label{figs:end_to_end}
        \end{figure}

        Fig.~\ref{figs:routing_eval} shows the accuracy of correct tool routing per classes in a confusion matrix. With an F1-score of 0.77 at the macro level, we could confirm high accuracy for each class except for RAG+SQL+PRD. The top-level model is capable of reasoning about the necessary steps to fulfill the query with proper delegation. Further, Fig.~\ref{figs:end_to_end} shows the quality of the synthesized result from the 181 examples evaluated by subject matter experts using the Likert scale. With a mean score of 3.89 overall, we confirm EPIC can return acceptable results except for classes that involve predictions (PRD, SQL+PRD, RAG+SQL+PRD).

        \textbf{Finding 6: }\textit{The top-level query processing module leveraging proprietary foundation models are capable of handling complex analysis questions by understanding the steps to fulfill the user query in an autonomous fashion.}

    \subsection{Inference Cost Evaluation}
        Advanced LLM models, can incur significant costs when running inference tasks. Therefore, cost evaluation is a critical component when designing and implementing LLMs. 
        We compare the cost of inference on the \textit{Llama-3-8B} model when hosted by a cloud service provider and accessed through a serverless API. 
        We use the number of tokens and the cost per token as metrics. 
        Tokens can be words, parts of words, or individual characters, and language models break down text into these tokens to understand and generate text.
        We compare the costs of serverless APIs from OpenAI and the Llama-3 8B model. The per-token cost for OpenAI is \$0.0000025 for input tokens and \$0.000010 for output tokens. For the Llama-3 8B model, the per-token cost is \$0.000003 for input tokens and \$0.0000061 for output tokens.
        
\begin{figure}[t]
              \centering
              \includegraphics[width=\linewidth]{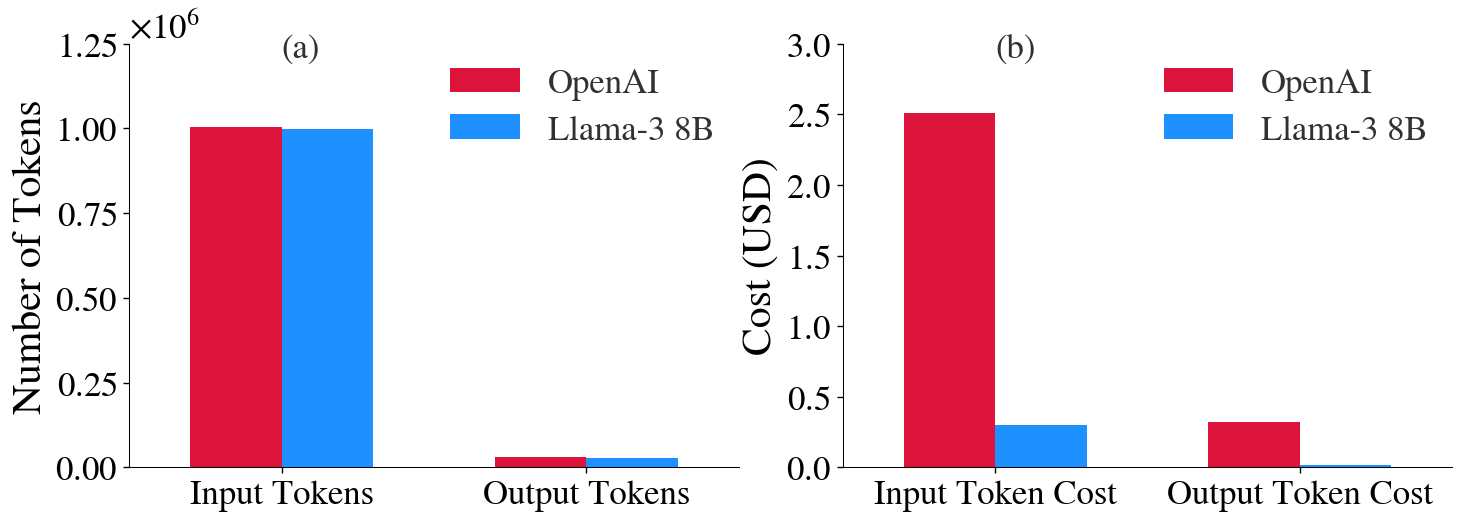}
              \caption{Figure illustrating comparison of token count \& inference cost of OpenAI and Llama models.}
              \label{figs:cost_eval_tokens}
        \end{figure}

        In Figure~\ref{figs:cost_eval_tokens}(a), we show the total number of tokens used by all the questions in \textit{DA} module evaluation tasks. Input tokens are calculated from queries asked to the models and output tokens are generated by response of the models. We observe that, for both models have more input token than output token. Also number of tokens of OpenAI model is similar to Llama models. However, the cost of using these tokens is significantly higher for OpenAI models (shown in Figure~\ref{figs:cost_eval_tokens}(b)). \\
        \textbf{Finding 7: } \textit{The cost of operating Llama model is 8.3 times less than that of OpenAI-4o models for input tokens and likewise, for output tokens, cost of Llama model is  19.1 times less compared to OpenAI-4o model. Thus, we demonstrate that the relatively smaller Llama models can do better or comparable in terms of performance evaluation at a fraction of the cost.}  
    
\section{Conclusion}
\label{sec:conclusion}
With the increasing operational complexity, ODA has become one of the key components of achieving high operational efficiency in HPC. However, it has hardly been a trivial endeavor for HPC centers to make impact with data. There is an unmistakable gap between data and insight mainly bottlenecked by the complexity of handling large amounts of data from various sources while fulfilling multi-faceted use cases targeting many different audiences. We observe that the ODA is bottlenecked by manual data exploration efforts, campaigns, analysis, synthesis, interpretation, and visualization of data.
    
To solve this “last mile” issue of data, we have presented the design and implementation of EPIC, a new capability that accelerates data exploration and analysis tasks in HPC ODA systems by leveraging the emerging capabilities of generative AI. 
EPIC’s dynamic, modular design enables it to adapt to evolving analytics tasks. 
We have found that the emerging reasoning and synthesis capabilities large language models can automate the synthesis multiple modalities of operational data; 1) unstructured HPC domain knowledge – text, image, and tables, 2) structured traditional tabular data, and 3) consumption of models of data for predictions.  Further, we have found that such reasoning and synthesis can be integrated in existing HPC ODA systems in a cost-effective, modular and extendible way by integrating generative AI by leveraging multiple agentic loops collaborating in a hierarchical way.

Through extensive evaluations, we have demonstrated that these capabilities are effective in handling each modality handling various ODA tasks. Also, these tasks would be reasoned about and be executed in an automated workflow to fulfill complex queries.
As future work, we aim extend this system to handle a wider array of tasks by introducing more lower-level capabilities better integrated with the real-time nature of ODA systems. Also, we aim to further explore and enhance the trustworthiness and reliability of end-to-end agentic approaches by tackling the problem of system evaluations.

\bibliographystyle{IEEEtran}
\bibliography{EPIC_25}

\end{document}